\documentclass[twocolumn,showpacs,preprintnumbers,amsmath,amssymb]{revtex4}
\usepackage{graphicx}
\usepackage{dcolumn}
\usepackage{bm}
\usepackage{amsmath, amsthm, amssymb}

\begin{document}

\title{Dissipation and Decoherence in Nanodevices: a Generalized Fermi's Golden Rule}

\author{D. Taj}
\email{david.taj@gmail.com}

\author{R. C. Iotti}
\author{F. Rossi}

\affiliation{Dipartimento di Fisica, Politecnico di Torino,
Corso Duca degli Abruzzi 24, 10129 Torino, Italy}


\date{13 november 2008}

\begin{abstract}
We shall revisit the conventional adiabatic or Markov
approximation, which ---contrary to the semiclassical case--- does
not preserve the positive-definite character of the corresponding
density matrix, thus leading to highly non-physical results. To
overcome this serious limitation, originally pointed out and
partially solved by Davies and co-workers almost three decades
ago, we shall propose an alternative more general adiabatic
procedure, which (i) is physically justified under the same validity restrictions of the conventional Markov approach, (ii) in the semiclassical limit reduces to the standard Fermi's golden rule, and (iii) describes a genuine
Lindblad evolution, thus providing a reliable/robust treatment of
energy-dissipation and dephasing processes in electronic quantum
devices. Unlike standard master-equation formulations, the dependence of our approximation on the specific choice of the subsystem (that include the common partial trace reduction) does not threaten positivity, and quantum scattering rates are well defined even in case the subsystem is infinitely extended/has continuous spectrum.
\end{abstract}

\maketitle

Present-day technology pushes device dimensions toward limits
where the traditional semiclassical or Boltzmann theory~\cite{ST}
can no longer be applied, and more rigorous quantum-kinetic
approaches are imperative \cite{QT1,QT2}. However, in spite of the
quantum-mechanical nature of electron and photon dynamics in the
core region of typical solid-state nanodevices
---e.g., superlattices \cite{SL} and quantum-dot structures
\cite{QD1,QD2,QD3}--- the overall behavior of such quantum systems
is often governed by a complex interplay between phase coherence
and energy relaxation/dephasing \cite{RMP}, the latter being also
due to the presence of spatial boundaries \cite{Frensley}.
Therefore, a proper treatment of such novel nanoscale devices
requires a theoretical modeling able to properly account for both
coherent and incoherent ---i.e., phase-breaking--- processes on
the same footing.

The wide family of so-called solid-state quantum devices can be
schematically divided into two main classes: (i) a first one which
comprises low-dimensional nanostructures whose electro-optical
response may be safely treated within the semiclassical picture
\cite{SP} (e.g., quantum-cascade lasers \cite{QCL1,QCL2,QCL3}),
and (ii) a second one grouping solid-state devices characterized
by a genuine quantum-mechanical behavior of their electronic
subsystem (e.g., solid-state quantum logic gates \cite{QLG1,QLG2})
whose quantum evolution is only weakly disturbed by decoherence
processes.

For purely atomic and/or photonic quantum logic gates, decoherence
phenomena are successfully described via adiabatic-decoupling
procedures \cite{QO} in terms of extremely simplified models via
phenomenological parameters; within such  effective treatments,
the main goal/requirement is to identify a suitable form of the
Liouville superoperator, able to ensure/maintain the
positive-definite character of the corresponding density-matrix
operator \cite{QOS}. This is usually accomplished by identifying
proper Lindblad-like decoherence superoperators
\cite{QOS,Lindblad}, expressed in terms of a few crucial
system-environment coupling parameters \cite{constrains}.

In contrast, solid-state devices are often characterized by a
complex many-electron quantum evolution, resulting in a
non-trivial interplay between coherent dynamics and
energy-relaxation/decoherence processes; it follows that for a
quantitative description of such coherence/dissipation coupling
the latter need to be treated via fully microscopic models.

To this aim, motivated by the power and flexibility of the
semiclassical kinetic theory \cite{ST} in describing a large
variety of interaction mechanisms, a quantum generalization of the
standard Boltzmann collision operator has been proposed
\cite{RMP}; the latter, obtained via the conventional Markov
limit, describes the evolution of the reduced density matrix in
terms of in- and out-scattering superoperators. However, contrary
to the semiclassical case, such collision superoperator does not
preserve the positive-definite character of the density-matrix
operator.

This serious limitation was originally pointed out by Spohn and
co-workers~\cite{D-S} three decades ago; in particular, they
clearly pointed out that the choice of the adiabatic decoupling
strategy is definitely not unique (see below), and only one among
the available possibilities, developed by Davies in~\cite{Davies}, could be shown to preserve positivity: it
was the case of a "small" subsystem of interest interacting with
a thermal environment, and selected through a partial trace
reduction. Unfortunately, the theory was restricted to finite-dimensional subsystems only (i.e., $N$-level atoms), and to the particular projection scheme of the partial trace.

Inspired by the pioneering papers by Davies and co-workers, aim of
the present Letter is to propose an alternative and more general
adiabatic procedure which (i) in the discrete-spectrum case
reduces to Davies' model~\cite{Davies}, (ii) for diagonal states
gives the well known Fermi's golden rule \cite{Alicki,FGR}, and
(iii) describes a genuine Lindblad evolution (depending on the relevant subsystem parameters), even in the infinite-dimensional/continuous spectrum case, thus providing a reliable/robust
treatment of energy-dissipation and dephasing processes in
semiconductor quantum devices.

Clearly, through our adiabatic-decoupling approach, different markovian approximations are generated by choosing different projection schemes (see below). However we stress that, contrary to standard master-equation formulations~\cite{Davies,MED}, these approximations are always of Lindblad type~\cite{Lindblad}, so that positivity is intrinsic in our adiabatic-decoupling strategy, and does not depend on the chosen subsystem.

Moreover, our approximation scheme holds true under the same validity regime of the conventional Markov approach: the so called weak-coupling limit~\cite{Davies2}, where the subsystem density matrix in the interaction frame moves slowly with respect of perturbative effects.

In order to recall the main features of the problem, let us
consider its general formulation based on the fully operatorial
approach proposed in \cite{PRB}. Given a generic physical
observable $A$ ---described by the operator ${\hat A}$--- its
quantum plus statistical average value is given by $A = {\rm
tr}\left\{{\hat A} {\hat \rho}\right\}$, where ${\hat \rho}$ is
the so-called density-matrix operator. Its time evolution is
dictated by the total (system plus environment) Hamiltonian, that
can be regarded as the sum of a noninteracting (system plus
environment) contribution plus a system-environment coupling term:
${\hat H} = {\hat H}_\circ + {\hat H}'$; the corresponding
equation of motion for the density-matrix operator ---also known
as Liouville-von Neumann equation--- in the interaction picture is
given by:
\begin{equation}\label{LvN_i}
{d{\hat \rho}^i \over dt}(t) = -i \left[\hat{\cal H}^i(t), {\hat
\rho}^i\right]\ ,
\end{equation}
where $\hat{\cal H}^i$ denotes the interaction Hamiltonian
$\hat{H}'$ written in units of $\hbar$.

The key idea beyond any perturbation approach is that the effect
of the interaction Hamiltonian ${\hat H}'$ is ``small'' compared
to the free evolution dictated by the noninteracting Hamiltonian
${\hat H}_\circ$. Following this spirit, by formally integrating
Eq.~(\ref{LvN_i}) from $t_\circ$ to the current time $t$, and
inserting such formal solution for ${\hat \rho}^i(t)$ on the
right-hand side of Eq.~(\ref{LvN_i}), we obtain an
integro-differential equation of the form:
\begin{equation}\label{IDE}
{d {\hat \rho}^i\over dt}(t)\! =\! -i \left[\hat{\cal H}^i(t),
{\hat \rho}^i(t_\circ)\right]\! -\! \int_{t_\circ}^t\! dt'\!
\left[\hat{\cal H}^i(t), \!\left[\hat{\cal H}^i(t'), {\hat
\rho}^i(t')\right]\right].
\end{equation}
We stress that so far no approximation has been introduced:
Equations (\ref{LvN_i}) and (\ref{IDE}) are fully equivalent, we
have just isolated the first-order contribution from the full time
evolution in Eq.~(\ref{LvN_i}).

According to the standard procedure, at this
point one should identify a subsystem of interest by, for example,
the use of a suitable projection on the space of density matrices~\cite{Davies,Davies2}:
although this step is crucial to study the weak-coupling limit, it
turns out, as we shall see, that it needs to be considered
explicitly only later on, when talking about positivity. For the
moment, we shall limit ourselves to use the projection $P_0$, to say that no first-order
terms are considered in this paper, i.e., we shall assume that
$P_0\left([\hat H'(t),P_0 {\hat \rho}]\right)= 0$. This assumption holds true and is physically justified for a very general form of the perturbing hamiltonian~\cite{Davies}, and allows us to neglect the first order term on the right hand side of (\ref{IDE}), as it gives no contribution when projected. In any case, one could always assume that condition above holds true, by assigning the first order quantity $P_0(\hat H')$ to the deterministic part of the hamiltonian $\hat H_\circ$ (which however could then be more difficult to solve).

Hence for ease of exposition we shall now study all the projection-independent
features of our model, safely neglecting first order terms.

Let us now focus on the time integral in Eq.~(\ref{IDE}). Here,
the two quantities to be integrated over $t'$ are the interaction
Hamiltonian $\hat{\cal H}^i$ and the density-matrix operator
${\hat \rho}^i$. In the spirit of the perturbation approach
previously recalled, the time variation of ${\hat \rho}^i$ can be
considered adiabatically slow compared to that of the Hamiltonian
$\hat{\cal H}$ written in the interaction picture, i.e.,
$\hat{\cal H}^i(t') = {\hat U}^\dagger_\circ(t') \hat{\cal H}
{\hat U}^{ }_\circ(t')$; indeed, the latter exhibits rapid
oscillations due to the noninteracting evolution operator ${\hat
U}_\circ(t) = e^{-{i{\hat H}_\circ t \over \hbar}}$. Therefore, in
the standard (and problematic) Markov approximation the
density-matrix operator ${\hat \rho}^i$ in interaction frame is simply taken out of the
time integral and evaluated at the current time $t$.

Following such prescription, the second-order contribution to the system dynamics
written in the Schr\"odinger picture
for the case of a time-independent interaction Hamiltonian $\hat{\cal H}$ comes out to be:
\begin{equation}\label{LvN-eff}
{d{\hat \rho} \over dt} =
-{1 \over 2} \left[\hat{\cal H},
\left[\hat{\cal K},{\hat \rho}\right]\right]
\end{equation}
with
\begin{equation}\label{calK}
\hat{\cal K}
=
2 \int_{t-t_\circ}^{0} dt'
\hat{\cal H}^i(t')
=
2 \int_{t-t_\circ}^{0} dt' {\hat U}^{\dagger}_\circ(t') \hat{\cal
H} {\hat U}_\circ(t') \ .
\end{equation}
The effective equation in (\ref{LvN-eff}) has still the double-commutator structure in (\ref{IDE}) but it is now local in time.
The Markov limit recalled so far leads to significant modifications
to the system dynamics: while the exact quantum-mechanical
evolution in (\ref{LvN_i}) corresponds to a fully reversible and
isoentropic unitary transformation, the instantaneous
double-commutator structure in (\ref{LvN-eff}) describes, in
general, a non-reversible (i.e., non unitary) dynamics characterized by energy dissipation and dephasing.
However, since any effective Liouville superoperator should describe the time evolution of
$\hat\rho$ correctly and since the latter, by definition, needs to be trace-invariant and positive-definite at any time,
it is imperative to determine if the Markov superoperator
in (\ref{LvN-eff}) fulfills this two basic requirements.
As far as the first issue is concerned, in view of its commutator structure, it is easy to show that this effective superoperator is indeed trace-preserving.
In contrast, as discussed extensively in \cite{PRB}, the latter does
not ensure that for any initial condition the density-matrix
operator will be positive-definite at any time.
This is by far the most severe limitation of the conventional Markov approximation.

By denoting with $\{\vert \lambda \rangle\}$, and neglecting energy-renormalization contributions~\cite{PRB}, the eigenstates of the noninteracting Hamiltonian $\hat{H}_\circ$, the effective equation (\ref{LvN-eff}) written in this basis is of the form:
\begin{equation}\label{LvN-eff-lambda}
{d\rho_{\lambda_1\lambda_2} \over dt} =
{1 \over 2} \sum_{\lambda'_1\lambda'_2}
\left[{\cal P}_{\lambda_1\lambda_2,\lambda'_1\lambda'_2}
\rho_{\lambda'_1\lambda'_2}
-
{\cal P}_{\lambda_1\lambda'_2,\lambda'_1\lambda'_1}
\rho_{\lambda'_2\lambda_2} \right] + {\rm H.c.}
\end{equation}
with generalized scattering rates given by:
\begin{equation}\label{calP}
{\cal P}_{\lambda_1\lambda_2,\lambda'_1\lambda'_2} = {2\pi \over \hbar} H'_{\lambda_1\lambda'_1} H^{\prime *}_{\lambda_2\lambda'_2} \delta(\epsilon_{\lambda_2} - \epsilon_{\lambda'_2}) \ ,
\end{equation}
$\epsilon_\lambda$ denoting the energy corresponding to the noninteracting state $\vert \lambda \rangle$. As discussed extensively in \cite{PRB}, such generalized scattering rates are obtained within the completed-collision limit, i.e., $t_\circ \to -\infty$, and neglecting energy-renormalization contributions.

The well-known semiclassical or Boltzmann theory~\cite{ST} can be easily derived from the quantum-transport
formulation presented so far, by introducing the so-called
diagonal or semiclassical approximation. The latter corresponds to
neglecting all non-diagonal density-matrix elements (and therefore
any quantum-mechanical phase coherence between the generic states
$\lambda_1$ and $\lambda_2$), i.e.,
$\rho_{\lambda_1\lambda_2} = f_{\lambda_1} \delta_{\lambda_1\lambda_2}$,
 where the diagonal elements $f_\lambda$ describe the semiclassical
distribution function over our noninteracting basis states.
Within such approximation scheme, the quantum-transport equation (\ref{LvN-eff-lambda}) reduces to the well-known Boltzmann equation:
\begin{equation}\label{BTE}
{d f_\lambda \over dt} =
\sum_{\lambda'} \left(
P_{\lambda\lambda'} f_{\lambda'} - P_{\lambda'\lambda} f_\lambda
\right)\ ,
\end{equation}
where
\begin{equation}\label{P}
P_{\lambda\lambda'} = {\cal P}_{\lambda\lambda,\lambda'\lambda'} =
{2\pi \over \hbar} |H'_{\lambda\lambda'}|^2 \delta\left(\epsilon_{\lambda}-\epsilon_{\lambda'}\right)
\end{equation}
are the conventional semiclassical scattering rates given by the well-known
Fermi's golden rule \cite{FGR}.

At this point it is crucial to stress that, contrary to the non-diagonal density-matrix description
previously introduced, the Markov limit combined with the semiclassical or diagonal approximation ensures that at any time $t$ our semiclassical distribution function $f_\lambda$
is always positive-definite.
This explains the ``robustness'' of the Boltzmann transport equation (\ref{BTE}), and its extensive application in solid-state-device modeling as well as in many other areas, where quantum effects play a very minor role.
In contrast, in order to investigate genuine quantum-mechanical phenomena, the conventional Markov superoperator in (\ref{LvN-eff}) cannot be employed, since it does not preserve the positive-definite character of the density matrix $\rho_{\lambda_1\lambda_2}$.

In order to introduce our alternative formulation of the problem,
let us go back to the integro-differential equation (\ref{IDE}),
and let us consider the following time symmetrization: given the
two times $t'$ and $t$, we shall introduce the ``average'' or
``macroscopic'' time $T = {t+t' \over 2}$ and the ``relative''
time $\tau = t-t'$. This change of variable has very solid bases, as it is common and well established in a wide variety of contests, such as Wigner's phase-space formulation of quantum mechanics~\cite{wigner}, standard quantum kinetics Green functions (see e.g.~\cite{haug}), and even classical radiation theory (e.g., in the treatment of Bremsstrahlung): the basic idea is that the relevant time characterizing/describing our effective system evolution is the
macroscopic time $T$. It is now easy to rewrite the second-order
contribution in Eq.~(\ref{IDE}) in terms of these new time
variables:
\begin{widetext}
\begin{equation}\label{IDE-new}
{d \over dT} {\hat \rho}^i(T) = - \int_0^{t-t_\circ} d\tau
\left[\hat{\cal H}^i\left(T+ {1 \over 2} \tau\right),
\left[\hat{\cal H}^i\left(T- {1 \over 2}\tau\right), {\hat
\rho}^i\left(T-{1 \over 2}\tau\right)\right]\right]\ .
\end{equation}
\end{widetext}
In the spirit of the adiabatic approximation previously recalled,
the density-matrix operator ${\hat \rho}^i$ can be taken out of
the time integral and evaluated at the current time $T$. As already stressed, this important approximation is valid when the density matrix varies slowly in interaction picture, that is, in the limit of weak-coupling. It is now convenient to replace the finite-domain time integration over $\tau$ by introducing a corresponding Gaussian correlation
function $e^{-{\tau^2\over 2 {\overline{t}}^2}}$ whose width
$\overline{t}$ may be regarded as a safe overestimation of the
so-called ``correlation time'', which is shorter than the period of the system evolution; indeed, for $t-t_\circ$ greater
than the correlation time, the time integration may be safely
extended up to infinity. To compare with the exact dynamics in the
weak-coupling limit, the correlation time has to finally be brought
back to infinity: this in turn may be accomplished by a scaling
property of the form
\begin{equation}\label{eq:asymptotic}
\overline t (g)\sim  g^{-\xi}\: \overline T,
\end{equation}
for $g\sim 0$, where $g$ is the coupling constant, $\overline
T$ is a fixed reference time, and $\xi>0$. So reformulating once again, we assume that the density matrix moves slowly with respect to the correlation time $\overline{t}$.

Focussing on the skew-adjoint part of Eq.~(\ref{IDE-new}), i.e.,
the so-called scattering part, which is clearly our main interest
(the self-adjoint part is just an energy-renormalization term and
does not threaten positivity), we get:
\begin{widetext}
\begin{equation}\label{LvN-eff-new1}
{d \over dT} {\hat \rho}^i(T) = -{ 1 \over 2 }
\int_{-\infty}^\infty d\tau\; e^{-{\tau^2\over 2
{\overline{t}}^2}} \left[\hat{\cal H}^i\left(T+ {1 \over 2}
\tau\right), \left[\hat{\cal H}^i\left(T- {1 \over 2}\tau\right),
{\hat \rho}^i\left(T\right)\right]\right]\ .
\end{equation}
\end{widetext}
We stress how the proposed time symmetrization gives rise to a
fully symmetric superoperator, compared to the strongly asymmetric
Markov superoperator in~\cite{Davies2}.

The second crucial step in order to get a genuine Lindblad
superoperator for the global dynamics is to exploit once again the
slowly-varying character of the density-matrix operator
$\hat\rho^i$ on the right-hand side of Eq.~(\ref{LvN-eff-new1}).
The key idea is to perform on both sides of
Eq.~(\ref{LvN-eff-new1}) a so-called temporal ``coarse graining'',
i.e., a weighted time average on a so-called microscopic scale, a
scale over which the variation of $\hat\rho^i(T)$ is negligible.
Since in the small and intermediate coupling regime such
time-scale is fully compatible with the correlation-time-scale $\overline{t}$, we shall perform such time average
employing once again a Gaussian correlation function of width
${\overline{t}\over 2}$, i.e.,
\begin{widetext}
\begin{equation}\label{LvN-eff-new3}
{d {\hat \rho}^i \over dT} (T) \!\!=\!\! -{1 \over
\sqrt{2\pi}\overline{t}}\! \int_{-\infty}^\infty\!\!\!\!\! dT'\: e^{-{2{T'}^2
\over \overline{t}^2}} \!\!\!\int_{-\infty}^\infty\!\!\!\!\! d\tau\: e^{-{\tau^2
\over 2\overline{t}^2}}\: \left[\hat{\cal H}^i\left(T\!+\!T'\!+\! {\tau \over
2} \right), \left[\hat{\cal H}^i\left(T\!+\!T'\!-\! {\tau \over
2}\right), {\hat \rho}^i\left(T\right)\right]\right]\ .
\end{equation}
\end{widetext}
Moving back to the original Schr\"odinger picture and combining
the two Gaussian distributions, the above equation can be
rewritten in the following compact form:
\begin{equation}\label{Lindblad-bis}
{d {\hat \rho}\over dT} = -{1 \over 2} \left[\hat{\cal L},
\left[\hat{\cal L}, {\hat \rho}\right]\right]
\end{equation}
with
\begin{equation}\label{calL-new}
\hat{\cal L} = \left({2 \over \pi\overline{t}^2}\right)^{1 \over
4} \int_{-\infty}^\infty dt' \;\hat{\cal H}^i(t')\; e^{-{{t'}^2
\over \overline{t}^2}} \ .
\end{equation}
This is the genuine Lindblad-like superoperator we were looking
for; indeed, the operators ${\cal L}$ are always Hermitian, and
such effective dynamics is positive-definite.

Let us finally rewrite the new Markov superoperator
(\ref{Lindblad-bis}) in our noninteracting basis $\lambda$,
defined by the (possibly generalized) eigenvectors of $H_\circ$:
we obtain an effective equation of motion of the form
\begin{equation}\label{LvN-eff-lambda}
{d\rho_{\lambda_1\lambda_2} \over dt}\! =\! {1 \over 2}
\sum_{\lambda'_1\lambda'_2}\! \left[{\cal
P}_{\lambda_1\lambda_2,\lambda'_1\lambda'_2}
\rho_{\lambda'_1\lambda'_2} \! -\!\! {\cal
P}_{\lambda_1\lambda'_2,\lambda'_1\lambda'_1}
\rho_{\lambda'_2\lambda_2} \right] \!+\! {\rm H.c.}
\end{equation}
with symmetrized quantum scattering rates
\begin{widetext}
\begin{equation}\label{calPtilde}
{\cal P}_{\lambda_1\lambda_2,\lambda'_1\lambda'_2} = {2\pi \over
\hbar} H'_{\lambda_1\lambda'_1} H^{\prime
*}_{\lambda_2\lambda'_2}\; {{1\over \sqrt{2\pi} \overline\epsilon}
\exp{\left\{-{
\left(\epsilon_{\lambda_1}-\epsilon_{\lambda'_1}\right)^2 +
\left(\epsilon_{\lambda_2}-\epsilon_{\lambda'_2}\right)^2 \over 4
\overline{\epsilon}^2 } \right\}} }
\end{equation}
\end{widetext}
substituting the strongly asymmetrical scattering superoperator given by the conventional Markov approximation (\ref{calP}).
Here, $\hbar$ has been shown explicitly, and $\overline{\epsilon}
= {\hbar\over\overline{t}}$ is a measure of the energy uncertainty
in the interaction process induced by our temporal coarse
graining. The above scattering superoperator can be regarded as
a generalization of the conventional Fermi's
golden rule to the density matrix formalism; indeed, in the semiclassical diagonal case
($\lambda_1=\lambda_2,\lambda_1'=\lambda_2'$) the above scattering superoperator boils down to what could be considered a dressed vertex-smoothed version of the Fermi's Golden Rule
\begin{equation}
P_{\lambda\lambda'} = {\cal P}_{\lambda\lambda,\lambda'\lambda'} = {2\pi \over
\hbar} |H'_{\lambda\lambda'}|^2\; {{1\over \sqrt{2\pi} \overline\epsilon}
\exp{\left\{-{
\left(\epsilon_{\lambda}-\epsilon_{\lambda'}\right)^2  \over 2
\overline{\epsilon}^2 } \right\}} } .
\end{equation}
and in the limit of infinite correlation-time ($\overline{\epsilon} \to 0$) the
standard scattering rates given by the Fermi's Golden Rule (\ref{P}) are readily recovered. It may be argued that the approximate dynamics thus found suffers from lack of generality, the gaussian smoothing having been put by hand. However, one has to look at asymptotic features only, as the approximation is valid in the weak-coupling limit $\epsilon\rightarrow 0$ only: in that limit, one does not appreciate the gaussian smoothing, but rather the (much more universal) asymptotic character of the correlation time (\ref{eq:asymptotic}). As such, we could say that our gaussian smoothing is a good representative among all the possible asymptotic markovian approximations to the exact dynamics, that guarantee a positive evolution.

In passing, we note that the transition rates (\ref{calPtilde}) could be regarded as a ``Quantum'' version of the celebrated Fermi's Golden rule. This should not generate confusion: of course the Fermi's Golden Rule transition rates are computed according to quantum mechanical calculations (see~\cite{cohen} among thousands of textbooks), but, once computed, they give rise to the Boltzmann equation (\ref{BTE}), which describes a \emph{classical} Markov process~\cite{ikeda} for \emph{classical} probabilities. Instead, the transition rates (\ref{calPtilde}) do \emph{not} describe a classical Markov process, but rather its \emph{quantum} analog: a so called Quantum Dynamical Semigroup~\cite{Lindblad} for the full density matrix.

Moreover, we would also like to note that there are many proposed Quantum generalization of the well known Fermi's Golden Rule (see e.g.~\cite{open}), which are robust, physically and mathematically meaningful. However, all these generalizations consider a bipartite system, one of which is finally traced over, the other being an $N$-level atom (a rather particular case). On the contrary, our model is much more general and refers to a closed quantum system, similarly to the original idea by Fermi~\cite{FGR}.

As pointed out previously, the whole theoretical scheme becomes
meaningful and applicable only when a well-defined subsystem of
interest is identified (together with a corresponding
infinite-dimensional environment), so that its correlation
time $\overline t$ can be estimated, and our (irreversible)
semigroup dynamics can correctly describe the projected (but fully
reversible~\cite{QOS}) exact Hamiltonian dynamics. As final crucial step, we
shall show that our conclusions about positivity remain valid no
matter how the subsystem is chosen.

To this end, we notice that the usual partial-trace projection,
when viewed in Heisenberg picture, is of the form $P_0:\sum_n \hat
A_n\otimes \hat B_n \mapsto \left(\sum_n \textit{Tr}(\hat \omega
\hat B_n) \hat A_n\right)\otimes 1$, where $\hat\omega$ is the
"environment" density matrix, $\hat A_n$ and $\hat B_n$ are,
respectively, generic system and environment observables. From
this structure, it follows that the partial trace is completely
positive (see e.g.~\cite{Lindblad}). Moreover, the projected
observables, all being of the form $\hat A\otimes 1$, constitute a
subalgebra of the global-observable algebra. Based upon these two
key remarks, let us now consider a generic projection $P_0$ on a
subalgebra $\mathcal X$ of the space of observables in Heisenberg
picture, which is also a completely positive map, and let $\{\hat
V_\alpha\}$ be its Kraus decomposition~\cite{Kraus}, so that $P_0
\hat A = \sum_\alpha \hat V_\alpha^\dagger \hat A \hat V_\alpha$.
Then one could easily verify that $\mathcal{X}$ is made by
observables that commute with each of the $\hat V_\alpha$ and
$\hat V_\alpha^\dagger$. We now observe that, due to its symmetry,
our generator keeps the same form of eq.(\ref{Lindblad-bis}) also
in Heisenberg picture: by projecting the latter with $P_0$, and
using the completeness relation $\sum_\alpha \hat V_\alpha^\dagger
\hat V_\alpha=1$, one can easily write the form for the
subsystem's generator in Schr\"odinger picture, dual to the
projected dynamics on the subalgebra $\mathcal{X}$:
\begin{equation}
{d\over dT} {\hat \rho}=-{1\over 2}\sum_{\alpha\beta} \{{\hat
D_{\alpha\beta}}^\dagger \hat D_{\alpha\beta},{\hat\rho}\}
+\sum_{\alpha\beta} {\hat D_{\alpha\beta}} {\hat\rho} \hat
D_{\alpha\beta}^\dagger.
\end{equation}
Here the "quantum transition amplitude"
operators are given, according to (\ref{calL-new}), by
\begin{equation}
\hat{D}_{\alpha\beta}=\hat{V}_\alpha \hat{\mathcal{L}} \hat{V}_\beta
\end{equation}
(note that these operators are indeed $\overline t$-dependent, as can be seen from (\ref{calL-new})).
This Lindblad form \cite{Lindblad} shows indeed that we have
obtained the generator of the completely positive Quantum
Dynamical Semigroup we were looking for. Moreover, the effort we
made to consider this rather abstract class of projection is
completely justified, as, for example, it paves the way for a new
formalism for Quantum Transport: suppose that $\hat P$ is a
projection in our Hilbert space that identifies, say, a
one-dimensional nano-device, and let $\hat Q_l$ and $\hat Q_r$
project on the left and right contact respectively (obviously
$\hat P+\hat Q_l+\hat Q_r=1$). Then $P_0 \hat A=\hat P \hat A \hat
P + \hat Q_l \hat A \hat Q_l+\hat Q_r \hat A \hat Q_r$ does belong
to the class of projections we have just studied, but the chosen
subsystem does not come from a partial trace, nor does it have
finite dimensions or discrete spectral properties: its
weak-coupling dynamics needs the full power of our theory, in
contrast with the previous ones \cite{Davies,Davies2}.

At this point a few comments are in order. As discussed
extensively in \cite{PRB}, also for the simplest case of a
standard two-level system ---i.e., a generic quantum bit--- the
standard Markov superoperator predicts a non-trivial coupling
between level population and polarization described by the
so-called $T_3$ contributions. In contrast, for a two-level system
coupled to its environment, the proposed quantum Fermi's golden
rule does not predict any $T_3$ coupling term (they vanish in the
infinite correlation time limit $\epsilon\rightarrow 0$), thus providing
a rigorous derivation of the well-known and successfully employed
$T_1 T_2$ dephasing model \cite{SL}. In general in fact, one could
show that for subsystems with discrete spectra, the limit $\epsilon\rightarrow 0$ reduces to Davies'
theory~\cite{Davies}. However, for subsystems with continuous
spectra, such limit is not defined, but for all finite collision
times $\overline t >0$ the proposed approach gives $\overline t
\sim T_3$, so $T_3$ contributions are indeed present, but they
become less and less important as the collision time $\overline t$
is raised, as it must be in the weak-coupling limit, to compare
with the exact hamiltonian dynamics (see before).

To summarize, we have proposed a new approach to modeling nowadays semiconductor nanodevices, by critically reviewing the standard Markov procedure. Indeed, the latter does not preserve the positive-definite
character of the density-matrix operator, thus leading to highly
non-physical results. To overcome this serious limitation, we have
identified an alternative and more general adiabatic procedure
which (i) is physically justified under the same validity restrictions of the conventional Markov approach, (ii) in the semiclassical limit reduces to the standard
Fermi's golden rule, and (iii) describes a genuine Lindblad
evolution, thus providing a reliable/robust treatment of
energy-dissipation and dephasing in state-of-the-art quantum
devices. We stress that our formulation generalizes preexisting
theories significantly, as it gives a positive dynamics for a
considerably large class of projections, i.e. ways to chose the
subsystem, and it is well defined even for infinitely extended subsystems (i.e. with continuous spectrum). In turn, on one side this allows to investigate
subsystems with both discrete and continuous spectra, a feature
largely shared by mesoscale electronic and opto-electronic quantum
devices; on the other side, it suggests a new way to treat
electrical contacts for quantum devices, thus opening up the
exciting possibility of a new formalism for Quantum Transport.

\begin{acknowledgments} 
We acknowledge Prof. Hisao Fujita Yashima (Dept. Mathematics,
University of Turin) and Prof. Paolo Zanardi (University of Southern California) for many stimulating and fruitful discussions.
\end{acknowledgments}

\end{document}